\documentclass[oldversion]{aa} 
\usepackage{graphicx}
\usepackage{natbib}
\bibpunct{(}{)}{;}{a}{}{,} 

\begin{document}
\date{}

\title{Correlation of the  scattering and dispersion events in the Crab Nebula pulsar}
\author{
A. Kuzmin\inst{1}
\and 
B.~Ya. Losovsky\inst{1}
\and
C.~A.Jordan\inst{2}
\and
F.~Graham Smith\inst{2}}

\institute{
Pushchino Radio Astronomy Observatory of Lebedev Physical Institute, Russia
\and
University of Manchester,
Jodrell Bank Observatory, Macclesfield, Cheshire, SK11~9DL, UK}

\abstract{ In separate series of observations of the Crab pulsar, pulse 
broadening due to scattering was measured at 111 MHz, and variations
of dispersion due to pulse delay were measured at higher radio
frequencies. In a remarkable event lasting 200 days a large
increase occurred in both parameters and with similar time signatures.
The increases in scattering and dispersion measure observed over the
200 days MJD 53950 -- 54150 are attributable to the effects of an ionised
cloud or filament crossing the line of sight.  The cloud would be $10^{11}-
10^{12}$ m across, with electron density $10^3- 10^4$ cm$^{-3}$.  
The increased scattering might originate within the cloud itself, or the moving
filament might induce turbulence in a separate higher density cloud in
the line of sight.}

\keywords{pulsars: general --
          pulsars: individual: Crab --
          supernova remnants
          }

\maketitle

\section{Introduction}
It has been known for many years that radio pulses from the Crab
 pulsar are affected both by a variable delay due to changes in
 dispersion and by a variable pulse broadening due to scattering
 along the line of sight (\cite{rc73}; \cite{ir77}).  Both phenomena
 vary on a typical time scale of about 100 days, but in previous 
 observations their
 variations have appeared to be imperfectly correlated and
 possibly even uncorrelated. At our two observatories, we have
 maintained for several years two series of observations to
 separately monitor these two phenomena, and can now report a discrete
 event that shows a remarkably good correlation between variations in
 scattering and in dispersion measure.

\section{Observations}
Observations of dispersion measure are made at least once a week at
Jodrell Bank Observatory as part of the Crab pulsar timing ephemeris
which has been produced and made generally available since 1982. The
ephemeris is based on daily observations of time of arrival of pulses
at 610 MHz, while the dispersion delay is measured by comparison with
similar observations at 1400 MHz. Observations at Pushchino Radio
Astronomy Observatory monitoring the pulse shape at 111 MHz have
continued since 2004. Both before and during the event the pulse is
broadened with a steep rise and an approximately exponential decay
with a time constant of several milliseconds; this characteristic decay
time is monitored almost daily.

A distinctive property of the Crab pulsar low frequency observations
is that the scatter broadening may be comparable with or greater than
the pulsar period. To avoid the resulting confusion we use for
observations the giant pulses of this pulsar, which stand out of the
regular pulses as rare, strong, well defined single pulses. The pulse
broadening is measured by fitting the convolution of a Gaussian
template pulse with a truncated exponent as the thin screen scatter
function, to the observed pulsar pulse.

\begin{figure*}
\begin{center}
    \includegraphics  {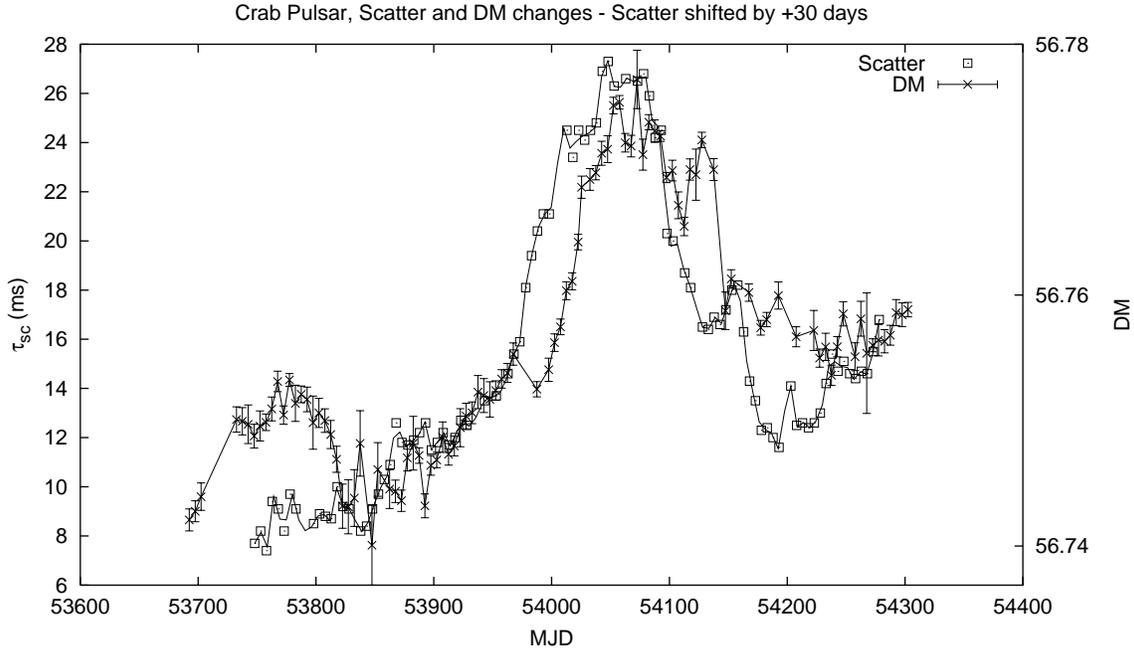} 
\caption {Pulse scatter length at 111 MHz, and dispersion measure DM.}
\end{center}
\label{fig:dmscat}
\end{figure*}

The results of these measurements over a period of 600 days are shown
in Figure \ref{fig:dmscat}. This shows a discrete event, lasting 200
days (MJD 53950 -- 54150), during which the dispersion and scattering
changed together.  The two curves are shown as recorded; note
especially the sharp rise at the start of the event, and the delay of
30 days between the onset of the rise in scattering and the rise in
DM. Both before and after this event there are smaller variations
which are less obviously correlated. The event appears as a distinct
phenomenon which stands out from the general level of variation in
both parameters.

The dispersion measure is proportional to the total electron content
along the line of sight. Most of this is attributed to the
interstellar medium, and this component is not expected to show such
large and rapid variations: observations of other pulsars show only
comparatively small and slow variations, as shown by \cite{you07}.
The base level of the dispersion measure appears to be 56.745 
cm$^{-3}$pc; the event increases this by 
$\Delta$DM $\approx$ 0.03 cm$^{-3}$pc. The observed scattering, 
by contrast, is more than
doubled at the event, increasing from 10 to 25 ms. Scattering by
random variations in refractive index depends on irregular
fluctuations of electron density within any part of the propagation
path; the simplest interpretation is that the increased scattering and
the increased dispersion are both due to a discrete electron cloud or
filament within the Nebula.
\section{Size and Location}  
Three length scales are involved in estimating the size of a single 
cloud responsible both for increased dispersion and scattering:

\begin{enumerate}
 \item The overall size $R$ of the cloud, which we will estimate from
 the duration of the event and the range of velocities which might be
 involved.  We will assume that the thickness is also $R$.  A maximum
 overall size may be estimated by assuming the velocity of the cloud
 across the line of sight does not exceed 100 km{\thinspace}s$^{-1}$, 
 which for a duration of 200 days gives a maximum dimension 
 $R \leq 2 \times 10^{12}${\thinspace}m ( $= 7 \times 10^{-5}${\thinspace}pc).

 \item The distance $L$ from the pulsar to the cloud, which may be
related to the observed wisps close to the pulsar (\cite{bhfb04}),
i.e. $L \approx$ 0.1 pc, or it may be related to structure at the
outer edge of the Crab Nebula (\cite{san+98}), i.e. $L \approx 1.5$
pc.

\item Scattering through the cloud appears as a disc whose radius $r$
is related to the pulse broadening.  This disc must be smaller than
the cloud itself; it must however be not much smaller, since
scattering was observed before the cloud was in the direct line of
sight. Since the observer is at a large distance compared with $L$,
the pulse broadening $\tau$ is simply related to the scattering angle
$\theta_{\rm rms}$ by $\tau = L\theta_{\rm rms}^2/2c$, giving a scattering
angle of 8 arcseconds for $L = 0.1$ pc and 2 arcseconds for $L = 1.5$
pc. The radius of the scattering disc is $L\theta_{\rm rms}$ , which is
$10^{11}${\thinspace}m for $L = 0.1$ pc and 
$4.5 \times 10^{11}${\thinspace}m for $L = 1.5$ pc. The cloud must
therefore have a minimum dimension $R \ge 10^{11}${\thinspace}m.

\end{enumerate}
      
Given these limits on $R$, and assuming the cloud has the same
dimension along the line of sight, the observed change $\Delta$DM in
dispersion measure gives an electron density $n_e = 10^3 - 10^4$
cm$^{-3}$. This high density is a strong indication that the cloud is
within the Nebula, and probably associated with activity near the
centre of the Nebula.  We note also that \cite{rcip88} observed a
change in Faraday rotation associated with a similar change in
dispersion, which they interpret as a magnetic field value of 160{\thinspace}
$\mu$G within a filament of ionised gas. These events may be related to
a larger and more complex pulse scattering event observed in 1997-8
(\cite{lpg01}).

The sharp onset of both scattering and increased DM indicate that the
leading edge of the cloud is well-defined, with a maximum scale of
about $3\times 10^{10}${\thinspace}m.  The earlier onset of scattering
is consistent with our model, since scattering is expected to be
observable before the direct line of sight is crossed by the cloud.

Other interpretations of this event, involving more than one component
of the Nebula, are of course possible.  In particular, we note that
\cite{klll08} have shown that a pulse time delay observed at low
frequencies, {\it additional to the frequency $\nu^{-2}$ dependence},
indicates the existence of a region in the line of sight with an
electron density as high as $10^6$cm$^{-3}$. Turbulence in the moving
cloud which we now observe may have been induced by a collision with
such a dense region.

Summary.   The increases in scattering and dispersion measure observed
over the 200 days MJD 53950-54150 are attributable to the effects of
an ionised cloud or filament crossing the line of sight.  The cloud
would be $10^{11}- 10^{12}$ m across, with electron density $10^3-
10^4$ cm$^{-3}$.  The increase in scattering might be due to the cloud
itself, or to turbulence induced in a high electron density cloud in
the line of sight.
\section{Acknowledgements}
Research at the Pushchino Radio Astronomy Observatory was partly
supported by Russian Foundation for Basic Research N
05-02-16415. A.Kuzmin and B.Losovsky thank A.Aleksandrov, V.Ivanova,
I.I.Litvinov, S.Logvinenko, T.Semina and M.Tchereshnev for assistance
in observations and data reduction.

\bibliographystyle{aa}
\bibliography{journals,psrrefs,modrefs,myrefs}

\end{document}